# Stokes Law at Molecular Length Scales: Effects of Intermolecular Interactions and Linear Response Theory

**Subhajit Acharya and Biman Bagchi***

Solid State and Structural Chemistry Unit

Indian Institute of Science, Bengaluru, India

*corresponding email: bbagchi@iisc.ac.in; profbiman@gmail.com

## Abstract

The celebrated Stokes Law (SL) of hydrodynamics predicts that the velocity of a particle pulled through a liquid by an external force, $F^{ex}$, is directly proportional to the force and inversely proportional to the friction $\zeta$ acted by the medium on the particle. We investigate the range of validity of Stokes Law at molecular length scales by employing computer simulations to calculate friction by pulling a tagged particle with a constant force. We thus calculate friction *for two model interaction potentials*, Lennard-Jones, and soft sphere, for several particle sizes, ranging from radius ($a$) smaller than the solvent particles to three times larger. We next obtain friction from diffusion ($D$) by using Einstein's relation between diffusion and friction $\zeta$ in an unperturbed liquid. We find, to our surprise, a quantitative agreement between the two at a small-to-intermediate pulling force regime for all the sizes studied. The Law does break down at a large pulling force beyond a threshold value. Importantly, the range of validity of Stokes' scheme to obtain friction increases substantially *if we turn off the attractive part of the interaction potential*. Additionally, we calculate the viscosity ($\eta$) of the unperturbed liquid and find a good agreement with the Stokes-Einstein relation $\zeta=C\eta a$ for the viscosity dependence with a value of $C$ close to $5\pi$, that is intermediate between the slip and stick boundary condition.

# I. INTRODUCTION

Stokes hydrodynamic relation, widely known as Stokes Law, provides a much-used relation between the time-independent, steady velocity $<v_{tagged}>$ of a tagged particle (as the response) when the particle is pulled at a constant force $F^{ex}$

$$<v_{tagged}> = \frac{F^{ex}}{\zeta} \qquad (1)$$

The absence of Newton's force-driven acceleration is due to dissipation, quantified by the friction term $\zeta$ that serves as the proportionality constant between $F^{ex}$ and $<v_{tagged}>$. The validity of Stokes Law and its extension as the Stokes-Einstein-relation between diffusion and viscosity have been extensively studied over the years and are even now a subject of much discussion.[1–3] Of particular interest to the physical chemistry/chemical physics community is the relation between diffusion and viscosity and the dependence of diffusion on the size of the moving solute.[4,5] Stokes-Einstein relation provides a simple relation where diffusion is inversely proportional to the product of viscosity ($\eta$) and radius ($a$) of the tagged particle. These relations can be tested in experiments.[6,7] Both theoretical and experimental studies have found evidence of the breakdown of Stokes-Einstein relations, and explanations have been offered.[8,9]

Stokes Law can be regarded as the first expression of the celebrated linear response theory that forms the cornerstone for both equilibrium and the time-dependent response of a system to an external perturbation.[10–12] While equilibrium response functions like specific heat, isothermal compressibility, etc., are determined by mean square fluctuations in the relevant quantities like energy and density, the time-dependent responses like friction, diffusion, conductivity, etc., are governed by the respective time correlation functions.[6,7,13,14]

Stokes Law for the velocity-force-friction relation ultimately provides, through the Stokes-Einstein relation, the much-needed connection between theory and experiments, such as diffusion and viscosity. Experiments routinely measure the viscosity dependence of the rate of a chemical reaction to understand the underlying molecular mechanism.[15,16] Einstein himself used Stokes relation to obtain the relation between diffusion coefficient and viscosity – the expression is provided below.[13] A common example is the study of the viscosity dependence of cis-trans isomerization in chemical and biophysical reactions.[17–20] These are activated processes and exhibit strong (inverse) viscosity dependence of the rate that plays a critical role in understanding the mechanism of the reaction.

In a pioneering work on the theory of Brownian motion, Einstein derived a relation between diffusion and friction[13]

$$D = \frac{k_B T}{\zeta} \tag{2}$$

Where $D$ denotes the diffusion coefficient of the tagged particle, $k_B$ is the Boltzmann constant, and $T$ is the absolute temperature. In a series of interesting works, Perrin verified Einstein's theory of diffusion by directly measuring the mean square displacement of the colloidal particles.[21] Perrin tracked the movement of the individual particles by projecting the light through the suspension of colloidal particles in a solution.[22]

Despite the success of Perrin experiments, one should remember that these experiments and hydrodynamic relations are restricted to the condition where the tagged particle is *much larger than the solvent (or bath) molecules*. One employs hydrodynamic boundary conditions to derive the relation between friction and viscosity, which cannot be quantified at a molecular level. Solvent molecules do not stick to the surface of the tagged molecule. Also, the tagged molecules cannot just slip past by without causing disturbances.[23] It is fair to state that despite the numerical success of the hydrodynamic expressions and extensive effort by theoreticians to extend the hydrodynamic approach to molecular scales, the approaches are fraught with approximations. An example in point is the failure of the ordinary Langevin equation to describe the time dependence of velocity time correlation function (TCF) of a solvent particle in liquid. The ordinary Langevin equation gives an exponential decay of the velocity TCF, while the measured decay has a rich non-exponential decay.[14] In a theoretical study, Zwanzig and Bixon[24] extended the hydrodynamic treatment of Stokes Law for a moving particle in a viscoelastic solvent with the use of frequency-dependent friction.[25] The study was motivated by the computer simulation results of the hard-sphere system by Alder and Wainwright that revealed a hydrodynamic-like velocity profile around the moving particle.[26] The derived expression provided a satisfactory explanation to understand the slow decay of velocity correlation caused by a vortex flow pattern in hard-sphere systems.[27]

There have been interesting recent developments in this field. In recent theoretical work, Squires and Brady considered a colloidal probe driven by an externally applied force in a suspension of neutrally buoyant bath particles.[28–30] to explain the shear thinning behavior of the effective viscosity in the presence of the external field. Later, Puertas and coworkers[31] studied the dynamics of a large tracer dragged with a constant small force (i.e., linear regime) in a bath of quasi-hard colloidal spheres with *Langevin dynamics simulation* and continuum mechanics. Their theoretical analysis was based on the Navier-Stokes equation, and the friction coefficient experienced by the probe showed *faster growth with the probe size than the prediction from Stokes Law* with both slip and stick boundary conditions.[32]

In a recent experiment, the friction coefficient was measured on a single molybdenum disulfide ($MoS_2$) nanotube using atomic force microscopy (AFM). *A nontrivial dependency of friction on interaction strength was revealed between the nanotube and the underlying substrate.*[33] In another simulation study, $MoS_2$ nanoparticles were subjected to different magnitudes of normal force.[34] The friction coefficient was calculated from the ensemble average of the ratio between the normal force and the shear force acting on the $MoS_2$ nanoparticle. It was observed that friction obtained from simulation agreed well with experimental measurements via the AFM study.

In this work, we ask the following specific questions:

(i) At what values precisely do the linear relation breaks down? Can we specify a threshold value for the pulling force, where the departure from the linear relation sets in? How small or large can this force be with respect to the root mean square fluctuation in force arising from natural interaction with the bath particles?

(ii) *How sensitive is this threshold force to the interaction potential between the tagged and the solvent molecules, beyond which we observe the deviation from the linearity*?

(iii) Does the threshold value of the force change with changing the thermodynamic state of the system?

(iv) Can we establish a microscopic connection that the increasing strength of intermolecular interaction causes an early departure from linearity?

(v) How does the size of the tagged particle affect the region of linearity in this context?

Towards answering the above questions, we employ molecular dynamics simulations to carry out "theoretical experiments" to obtain the drift velocity of the tagged particle in response to an applied force. We initially attempted pure molecular dynamics simulations in the microcanonical ensemble but observed the breakdown of Stokes law even at very small values of pulling force. We discovered that the failure happens due to the setup of non-hydrodynamic inertial flow in the liquid. We further found that adding a small random noise through the Langevin thermostat restores the hydrodynamic flow field. This small noise, quantified by a zero frequency friction, is uncorrelated with the systematic force from intermolecular interactions (that is, orthogonal to the systematic force) and thus poses no problem in subtracting it from the total friction calculated.

*What we found surprising is that the estimates from the direct execution of Stokes law are in complete quantitative agreement with that from Einstein's equation between diffusion and friction.* The second important result is the finding that when Stokes relation is combined with Einstein's relation (resulting in the well-known Stokes-Einstein viscosity-diffusion relation) with a value of the prefactor (usually denoted by "*C*"), it is exactly between the stick and the slip hydrodynamic boundary conditions and is in agreement with the mode coupling theory predictions of Keyes and Oppenheim, and Bhattacharyya and Bagchi.[35]

The rest of the paper is organized as follows. In sec-II, we briefly discuss the derivation of Stokes relation from linear response theory. Sec-III details the simulation protocol. In sec-IV, we examine the validity of Stokes Law for two different interaction potentials between the particles at two distinct thermodynamic state points. In sec-V, we investigate the diameter dependence of friction in the context of the Stokes-Einstein relation. In sec-VI, we attempt to understand the microscopic origin for the emergence of nonlinear response. Sec-VII summarizes our work with concluding remarks. Discussions on finite-size effects in the present problem, the velocity and density profiles are all contained in the Supplementary Material (SM), where we also discuss the connection with mode coupling theory. The latter is used to understand the microscopic origin of the emergence of nonlinear response.

## II.  STOKES RELATION FROM LINEAR RESPONSE THEORY

This section briefly discusses the derivation of Stokes relation by using the principles of linear response theory enacted by Kubo.[10] Here, we follow a procedure outlined by Zwanzig in his seminal work.[4] The method uses the perturbation Hamiltonian arising from the action of the external force term in the Liouville equation. Let us assume a tagged particle (say, #1) in a system consisting of $N$-1 bath particles is being pulled with a constant force $F^{ex}$ along the Z-direction from time $t=t_0$. Therefore, the total Hamiltonian of the system at time $t$ becomes

$$H_t = H_0 + H'_t \tag{3}$$

Where $H_0$ denotes the Hamiltonian of the system in the absence of force and is given by

$$H_0 = \sum_{i=1}^{N} \sum_{\alpha=x,y,z} \frac{(p_i^\alpha)^2}{2m} + U(q_1^x, ..., q_N^z) \tag{4}$$

Here $p_i^\alpha$ denotes the $\alpha$-component of the momentum of the $i^{th}$ particle, and $U$ is the potential energy of the system. In Eq.(3), the perturbative term $H'_t$ is given by

$$H'_t = -F^{ex}(t).q_1 = -\Theta(t-t_0)F^{ex}q_1^z \tag{5}$$

Here $F^{ex}(t) = \Theta(t-t_0)F^{ex}\hat{e}_z$, since a constant force, $F^{ex}$ is switched on along the Z-direction from time $t=t_0$. Here $\Theta$ is the Heaviside step function, $\hat{e}_z$ denotes the unit vector along the Z-direction, and $q_1^z(t)$ is the Z-coordinate of particle #1 at time $t$. The total Liouville operator can be written as

$$iL_t = iL_0 + iL'_t$$

Where $iL_0$ is given by

$$iL_0 = \sum_{i=1}^{N} \sum_{\alpha} \left[ \left(\frac{\partial H_0}{\partial p_i^\alpha}\right) \frac{\partial}{\partial q_i^\alpha} - \left(\frac{\partial H_0}{\partial q_i^\alpha}\right) \frac{\partial}{\partial p_i^\alpha} \right]$$
$$= \sum_{i=1}^{N} \sum_{\alpha} \left[ \frac{p_i^\alpha}{m} \frac{\partial}{\partial q_i^\alpha} + F_i^\alpha \frac{\partial}{\partial p_i^\alpha} \right] \tag{6}$$

In Eq.(6), $m$ is the mass of the particle, $F_i^\alpha$ denotes the $\alpha$-component of the force exerted on particle $i$. The additional term in the presence of external force is given by

$$iL'_t = \sum_{i=1}^{N}\sum_{\alpha}\left[\left(\frac{\partial H'_t}{\partial p_i^{\alpha}}\right)\frac{\partial}{\partial q_i^{\alpha}} - \left(\frac{\partial H'_t}{\partial q_i^{\alpha}}\right)\frac{\partial}{\partial p_i^{\alpha}}\right]$$
$$= \Theta(t-t_0)F^{ex}\frac{\partial}{\partial p_1^z} \qquad (7)$$

We start with the Liouville equation given by

$$\frac{\partial f(t)}{\partial t} = -iL_t f(t) \qquad (8)$$

Here, $f(t)$ is the density of microstates in the presence of the field at time $t$. In the presence of a weak field, the density of microstates $f(t)$ at time $t$ can be approximated as

$$f(t) = f_0 + \Delta f(t) \qquad (9)$$

Where $f_0$ denotes the density of microstates in the absence of the external field and $\Delta f(t)$ is the additional term arising due to the perturbation. We substitute Eq.(9) in Eq.(8) and simplify it to obtain[6]

$$\Delta f(t) = \frac{F^{ex}}{k_B T} f_0 \int_0^t ds\, e^{-iL_0(t-s)} v_1^z \qquad (10)$$

According to the linear response theory, the conjugate observable drift velocity at time $t$ can be simplified as

$$\begin{aligned}
\langle v_1^z(t)\rangle_t &= \int d\Gamma\, v_1^z \Delta f(t)\\
&= \frac{F^{ex}}{k_B T}\int d\Gamma f_0 \int_0^t ds\, v_1^z e^{-iL_0(t-s)} v_1^z\\
&= \frac{F^{ex}}{k_B T}\int d\Gamma f_0 \int_0^t ds\, \left(e^{iL_0(t-s)} v_1^z\right) v_1^z \qquad (11)\\
&= \frac{F^{ex}}{k_B T}\int d\Gamma f_0 \int_0^t ds\, v_1^z(t-s) v_1^z\;\; ;\text{since } v_1^z(t-s) = e^{iL_0(t-s)} v_1^z(0)\\
&= \frac{F^{ex}}{k_B T}\int_0^t ds\, \langle v_1^z(s) v_1^z(0)\rangle
\end{aligned}$$

In Eq.(11), $\langle v_1^z(t)\rangle_t$ denotes the nonequilibrium average of the Z-component velocity of the tagged particle (i.e., particle #1) at time $t$ in the presence of the external force. Here, we have employed the operator identity, i.e., $(iL_0)^\dagger = iL_0$ and a coordinate transformation for simplification in Eq.(11).[6] In the long-time limit (i.e., $t \to \infty$), Eq.(11) reduces as

$$\langle v_z(t) \rangle_{t \to \infty} = \frac{F^{ex}}{k_B T} \lim_{t \to \infty} \int_0^t ds \langle v_z(s) v_z(0) \rangle$$
$$= \frac{F^{ex}}{k_B T} D \qquad (12)$$

In Eq.(12), we have omitted the particle index and replaced $v_1^z$ as $v_z$ for simplification. According to LRT, the averaging needs to be taken over the initial conditions, and the initial state should be in the thermal equilibrium state without external perturbation. According to the Green-Kubo relation, the self-diffusion coefficient (D) is defined as $D = \frac{1}{d} \int_0^\infty d\tau \langle \mathbf{v_1}(\tau) \cdot \mathbf{v_1}(0) \rangle$; here, $d$ denotes the dimension of the system and $\mathbf{v_1}(\tau)$ is the velocity of the particle. Now, we employ Einstein's relation between diffusion and friction, i.e., $D = \frac{k_B T}{\zeta}$ to obtain

$$\langle v_z(t) \rangle = \frac{1}{\zeta} F^{ex} \qquad (13)$$

Eq.(13) is popularly known as Stokes Law, as described before. Einstein's relation connects diffusion with friction at the molecular scale, whereas Stokes relation provides a platform to estimate friction at the microscopic level.

It is important to note here that both the above derivation and the Stokes Law ignore terms higher than linear while deriving Eq.(13). However, in general, we can follow the cue from Kubo's linear response theory and attempt to perform a Taylor series expansion for the drift velocity of the tagged particle at time $t$ in terms of the external force $F^{ex}$ as follows.

$$\langle v_z(t) \rangle_t = \frac{1}{\zeta} F^{ex} + A(t)(F^{ex})^2 + B(t)(F^{ex})^3 + \ldots \qquad (14)$$

Here, $A(t)$ denotes the second-order derivative of the drift velocity to the external force, $B(t)$ is the third-order derivative, and so on. Since, at equilibrium, the average velocity of the tagged particle at time $t$ vanishes in the absence of the external, Eq.(14) does not contain any zeroth order term. When the strength of the external force is small, the first linear term primarily contributes over other order terms to determine the velocity of the tagged particle, and we recover Stokes law. However, as we increase the strength of the external force further, the higher-order terms would start to make thrift presence felt and ultimately dominate over the linear term, making the plot of drift velocity against the external force nonlinear. In this work, we monitor the threshold value of the external force, where the deviation from the linearity sets in the drift velocity plot against the external force for two different interaction potentials.

## III.  SIMULATION DETAILS

In our work, the system under study is composed of 10,976 particles in a cubic box with periodic conditions along *X*, *Y*, and *Z* directions. Let us assume the tagged particle *k* is pulled by a constant force $F^{ex}$ from time $t=0$ along the Z-direction. The motion of the tagged particle is governed by classical Newton's equation, while the remaining particles follow the ordinary Langevin equation. For the tagged particle *k* and solvent molecules, the equation of motion reads,

$$\dot{v}_k(t) = \frac{1}{m_k} \sum_{i \neq k} F_{ik} + \frac{1}{m_k} F^{ex} \hat{e}_z \tag{15.a}$$

$$\dot{v}_j(t) = \frac{1}{m_j} \sum_{i \neq j} F_{ij} - \zeta_{bare,j} v_j(t) + \frac{1}{m_j} R_j(t) \tag{15.b}$$

where $m_j$ denotes the mass of particle *j*, $F_{ij}$ is the interaction force between particle *i* and *j*, $\zeta_{bare,j}$ is the bare friction coefficient, $\hat{e}_z$ denotes the unit vector along the Z-direction, and $R_j(t)$ denotes the random or fluctuating force with $\langle R_j(t)\rangle = 0, \langle R_i(t)R_j(t')\rangle = 2\zeta_{bare}\delta_{ij}k_BT\delta(t-t')$. The strength of the random noise ($B$) is related to the friction force via the fluctuation-dissipation theorem, i.e., $B = \zeta_{bare}k_BT$, with $k_BT$ is the thermal energy. Eq.(15.a) is written considering the fact that particle #k is being pulled with a constant force $F^{ex}$, along the Z-direction. All the simulations are carried out in reduced units using LAMMPS software[36] with a timestep of $0.001\tau$, where $\tau = \sqrt{\frac{m\sigma^2}{\varepsilon}}$ is the unit of time with $m$ the mass, $\sigma$ the unit of length, and $\varepsilon$ the unit of energy.[37] In our calculations, we take $\sigma, \varepsilon$ and mass of the particles to be one. We perform the molecular dynamics simulations with two different values of bare friction (i.e., 0.5 and 1.0) at a reduced temperature of $T^*=0.8$ and a number density of $\rho^* = 0.7$. In the present context, the dynamics of the surrounding liquids are determined by Eq. (15.b). Essentially, the random noise term plays a crucial role in maintaining the kinetic temperature and, consequently, the thermodynamic state of the system. Since this is an external random force term, it is not correlated with the intermolecular force term acting on the tagged particle. Such a method has been implemented in several earlier studies.[29,38–40]

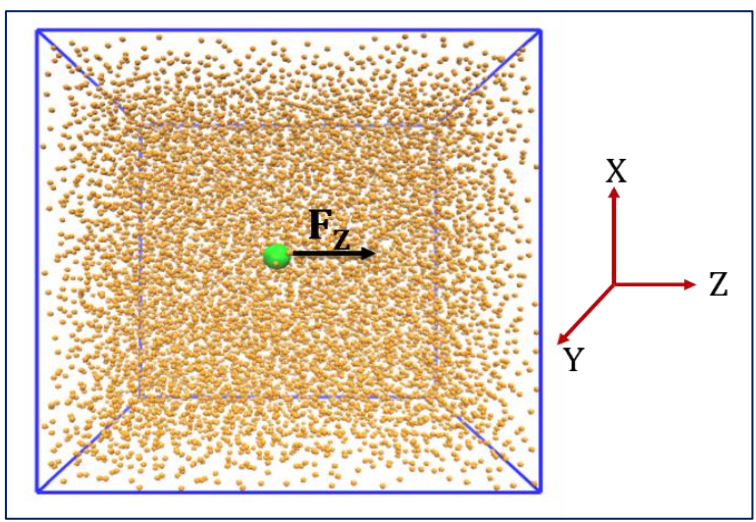

**Figure 1: Schematic diagram of the system simulated. We randomly choose one particle and pull it along the Z-direction with a constant force $F_Z$. We continue the pulling till a steady velocity is attained. We then use the average, steady velocity, and study Stokes Law.**

In order to understand the sensitivity of the interaction potential between the particles, we choose two different types of the interaction potential between particles $i$ and $j$

(a) $V_{LJ}(r) = 4\varepsilon\left[\left(\dfrac{\sigma}{r}\right)^{12} - \left(\dfrac{\sigma}{r}\right)^{6}\right]$

(b) $V_{SS}(r) = 4\varepsilon\left(\dfrac{\sigma}{r}\right)^{12}$ for $r \leq \sigma$
$\phantom{V_{SS}(r)} = 0;$ for $r > \sigma$

It is important to note that throughout the study, we maintain the strength of the fixed random noise to solely study the effects of the interaction potential on the emergence of nonlinear response. We randomly choose a single particle out of 10,976 particles and pull it with a constant force $F_Z$ along the Z-direction for $5 \times 10^6$ steps from time $t=0$, as shown in **Figure 1**, to calculate the steady-state velocity in response to a force that drags the tagged particle. In this study, we change the force over an extensive range and find that Stokes relation is valid over a wide range of force and velocity. We also vary the size of the tagged particle, keeping the size of solvent molecules intact to understand the effect of size on our calculations. We take the diameter of the tagged particle to be 0.5, 1.5, and 2.0 times that of the solvent molecules. To understand the sensitivity of the threshold force (where departure from Stokes linear relation sets in) to the thermodynamic state of the studied system, we also perform all the calculations at a different thermodynamic state point, i.e., $\rho^* = 0.85$ and $T^* = 1.0$.

## IV.   FORCE DEPENDENCE OF VELOCITY: EXAMINATION OF STOKES LAW

In this section, we discuss a multitude of results obtained mainly by different kinds of simulations at two different thermodynamic state points. We choose a single LJ particle out of 10,976 particles and pull it with a constant force $F_Z$ along the Z-direction from time $t=0$ for a long time until it reaches the steady state velocity. We vary the strength of the external force over a wide range and calculate the drift velocity of the tagged particle for each value of the applied force. It is to be noted that we consider the tagged particle to be the same size as the bath particles in these calculations. In **Figures 2(a)** and **2(b)**, we plot the average Z-component steady-state velocity of the tagged particle against the applied force $F_Z$ for two different values of bare friction, i.e., $\zeta^*_{bare} = 0.5$ and $\zeta^*_{bare} = 1.0$ (as shown by the black-colored filled circles). At the same time, in order to understand the effect of the interaction potential between the tagged particle and solvent molecules, we turn off the attractive part of the Lennard-Jones potential and carry out the same. In **Figures 2(c)** and **2(d)**, we plot the average Z-component steady-state velocity of the tagged particle against the applied force $F_Z$ for two different values of bare friction, i.e., $\zeta^*_{bare} = 0.5$ and $\zeta^*_{bare} = 1.0$ for Soft Sphere (SS) system (as shown by the black colored filled circles). To understand the sensitivity to the thermodynamic state of the system, we conduct the same to estimate the drift velocity of the tagged particle at the different thermodynamic state, i.e., $T^* = 1.0, \rho^* = 0.85$. **Figures 2(e)** and **2(f)** plot the drift velocity against the applied force for LJ and SS, as shown by the black-filled circles at bare friction $\zeta^*_{bare} = 1.0$, for this thermodynamic state (i.e., $T^* = 1.0, \rho^* = 0.85$). The vertical blue dotted line represents the threshold force beyond which nonlinear response emerges in **Figure 2**. We perform a linear fit in **Figure 2,** as shown by the red dashed line, to obtain Stokes Law friction using Eq.(13). We find that the linear relation works well in the low force regime; however, the deviation from the linearity sets in when the strength of the external force reaches the threshold value. We observe two significant trends: (i) the threshold value for the SS system is always higher than the LJ system, and (ii) the region of linearity gets reduced as we increase the density of the system. Due to the presence of an attractive part of the interaction potential between the particles, the cooperative effects that result in aligning the velocity vectors of the surrounding solvent molecules along the direction of pulling become prominent as we gradually increase the strength of the external force. However, the same effect is not prominent in the absence of an attractive part of the interaction potential. This is why nonlinear response emerges earlier in the LJ than that compared to the SS system. On the other hand, as we move on to different thermodynamic state points from T*=0.8, $\rho^* = 0.7$ to $T^* = 1.0, \rho^* = 0.85$, an early departure from the linearity in both LJ and soft sphere systems is observed, which can be attributed to the caging effect at higher density. Usually, in high-density liquid, the change in density dominates over the temperature change. In **Table 1**, we report the Stokes Law friction obtained by linear fitting in the case of the LJ and SS systems.

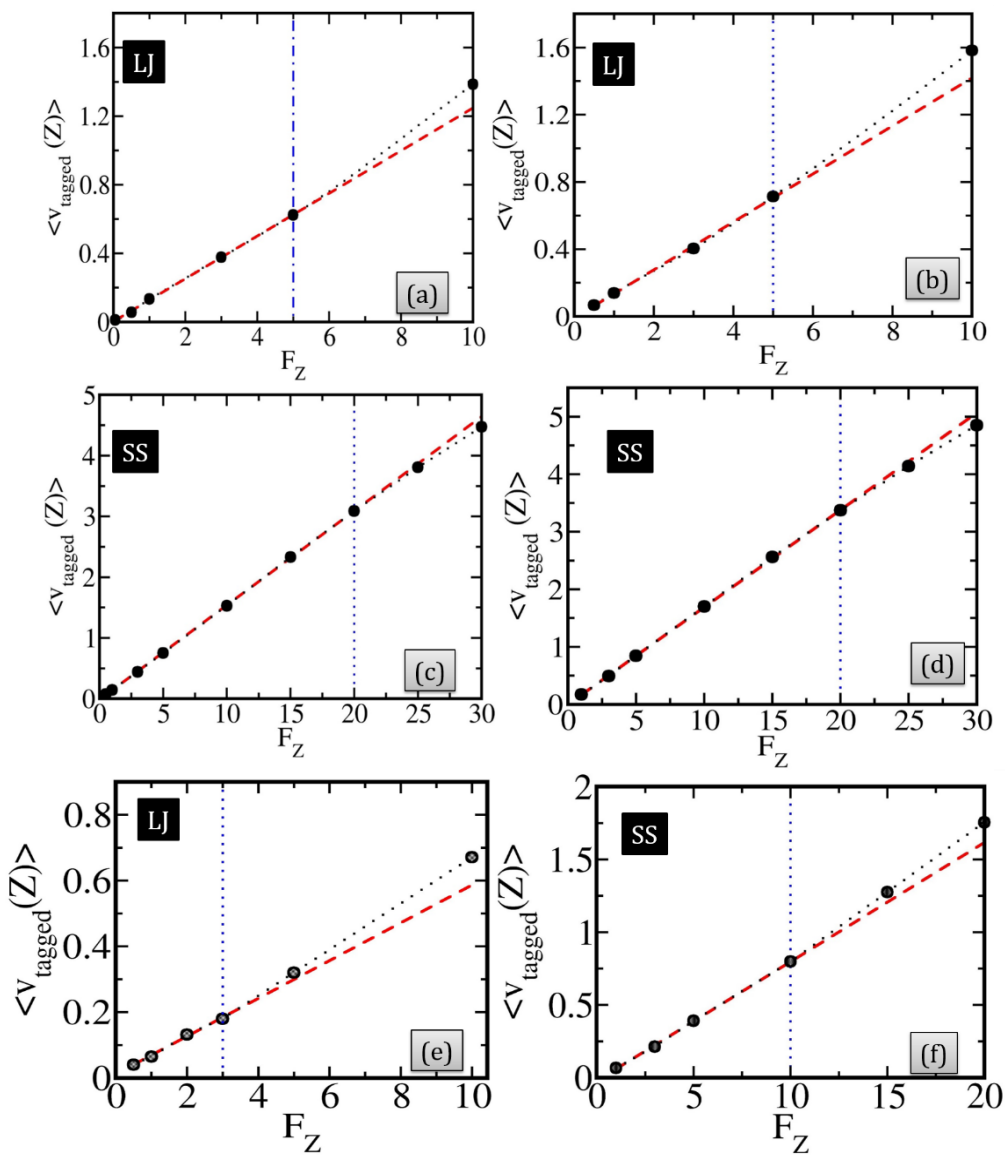

**Figure 2: The plot of the average Z-component steady-state velocity of the tagged particle against the applied force F$_Z$ in the Z-direction for three systems: (i) the LJ (*a* and *b*) (ii) soft sphere (*c* and *d*) system, and (iii) the same systems at higher density (*e* and *f*). In (a), the black colored circles show the variation of the average Z-component steady-state velocity of the tagged particle against the applied force at bare friction $\zeta^*_{bare} = 1.0$. A linear fit is performed to obtain the Stokes Law friction using Eq.(13), as shown by the red dashed line. We observe a deviation from the linearity when the external force exceeds 5.0, as shown by the vertical blue dotted line. In (b), we perform the same for bare friction $\zeta^*_{bare} = 0.5$. In (c), we validate the Stokes Law between force and drift velocity for the soft sphere system (SS) at bare friction $\zeta^*_{bare} = 1.0$. In this case, a deviation from the linearity is observed when the perturbative force is beyond 20.0, as shown by the vertical blue dotted line. In (d), we perform the same for bare friction $\zeta^*_{bare} = 0.5$ in the soft sphere system. In all cases, i.e.,(a-d), results are obtained from the molecular dynamics simulations carried out at T\*=0.8, and $\rho^* = 0.7$. We also study the same system at a different thermodynamic state (i.e., T\*=1.0, $\rho^* = 0.85$ and $\zeta^*_{bare} = 1.0$) and calculate the drift velocity of the tagged particle. In (e) and (f), we plot it against the applied force for LJ and SS, respectively. We observe that the region of linearity is reduced at this thermodynamic state in the case of both LJ and SS systems. In (a-f), the black dotted line connecting the black-colored data points is provided as a guide to the eyes. In all our calculations, we consider the probe particle to be the same size as the bath particles.**

On the other hand, we can calculate the friction from Einstein's relation between diffusion and friction. We compute the self-diffusion coefficient (*D*) of the tagged particle in the unperturbed liquid using Einstein's relation between *D* and the mean square displacement (MSD). In three dimensions, the self-diffusion coefficient is defined by[14]

$$D = \lim_{t \to \infty} \frac{\left\langle \left(r(t) - r(0)\right)^2 \right\rangle}{6t} \quad (16)$$

where *r*(t) is the position of the particle at time *t*, and angular brackets indicate the ensemble average. Alternatively, we can obtain the self-diffusion coefficient by integrating the un-normalized velocity autocorrelation. According to the Green-Kubo formalism, *D* is defined as,

$$D = \frac{1}{d} \int_0^\infty dt \langle \mathbf{v}(t) \cdot \mathbf{v}(0) \rangle \tag{17}$$

where $d$ indicates the dimension of the system and $\mathbf{v}(t)$ is the velocity vector of the particle at time $t$. Once we have diffusion from an equilibrium simulation, we can employ Einstein's relation between diffusion and friction, i.e., $D = \frac{k_B T}{\zeta}$ to estimate friction.

Einstein's method uses equilibrium MD simulation and evaluates friction from the diffusion constant $D$ via Einstein's relation, as reported in **Table 1**. On the other hand, Stokes Law calculates friction from a nonequilibrium simulation using the relation between force and drift velocity. It is necessary to stress that both methods involve a certain degree of approximation, so a study of the agreement between the two serves an essential purpose.

**Table 1** compares the two frictions obtained using Stokes Law and Einstein's method. We evaluate all these things for two different interaction potentials: (i) Lennard-Jones (6-12) potential and (ii) soft sphere $(1/r^{12})$ repulsive potential. We find surprisingly good agreement between the two entirely different methods for both systems.

**Table-1: Comparison between Stokes Law friction obtained from Figure 2 by fitting the linear regime of the plot of drift velocity against the applied force and from Einstein friction using Eq.(2).**

|  | *LJ system* |  | *Soft sphere* |  |
| --- | --- | --- | --- | --- |
| **Thermodynamic State** | *Stokes Law friction* | *Einstein friction* | *Stokes Law friction* | *Einstein friction* |
| $T^* = 0.8, \rho^* = 0.7, \zeta^*_{bare} = 0.5$ | 7.0±0.2 | 7.1±0.1 | 5.9±0.1 | 6.0±0.1 |
| $T^* = 0.8, \rho^* = 0.7, \zeta^*_{bare} = 1.0$ | 7.9±0.2 | 7.8±0.2 | 6.4±0.2 | 6.4±0.1 |
| $T^* = 1.0, \rho^* = 0.85, \zeta^*_{bare} = 1.0$ | 16.8±0.3 | 16.7±0.2 | 13.6±0.2 | 13.5±0.1 |

*The agreement between Stokes Law friction and Einstein friction is impressive and indeed surprising*, and at least to us, not fully expected for the following reasons. First, as mentioned above, Stokes Law is expected to be valid only when the tagged particle is much larger than the bath particles. That is, the tagged particle can be regarded as a Brownian particle. Second, we find a substantial presence of the inertial term in the decay of the velocity time correlation function. Of course, such contributions are already averaged in and included in the MSD estimates of diffusion. Yet, they do suggest the presence of microscopic effects. The velocity TCF decays exponentially in the Brownian limit, as given by the Langevin equation, without any external force term.[41] Thus, the averaging could be questionable for a small, tagged particle. However, the results demonstrated that there indeed lies hidden in some further understanding that we need to unearth.

## V. SIZE DEPENDENCE OF FRICTION FOR LJ AND SS INTERACTIONS

The Stokes-Einstein relation (based on Stokes Law and Stokes expression of friction) provides an expression for friction on a moving solute in terms of viscosity. This linear friction-viscosity relation has been widely used to understand many spectroscopic experiments, such as the rate of fluorescence quenching of aromatic molecules like Stilbene.[17,18,20,42–46] Here, the reactive motion involves the motion of a relatively bulky group around a body-fixed axis involving both rotational and translational motion. For translational friction, the relation is usually expressed as $\zeta = C\eta a$, where $\eta$ is the viscosity of the solvent, and $a$ is the radius of the tagged diffusing particle. The value of the constant "$C$" has been a subject of much discussion.

It is given as 6π for the stick hydrodynamic boundary condition and 4π for the slip boundary condition.[23]

Many experimental and theoretical results have found evidence for such a value of $C$ (i.e., between 4π and 6π) and also some that cast doubt on such simple viscosity dependence on friction [1,47–49]. Mode coupling theory analyses provide interesting results. These theories have consistently offered a value of $C$ between the stick and slip boundary conditions. A detailed mode coupling theory calculation by Bhattacharyya and Bagchi gave a value of $C=5.1\pi$ for the argon system near the triple point at T*=0.8 and $\rho^*=0.82$.[17,18] The numerical value of $C$ obtained from our calculation provides an understanding of the nature of the interaction between the tagged particle and the solvent molecules while pulling the tagged particle with a constant force in the case of LJ and SS systems. To estimate the numerical value of $C$, one must study the diameter-dependent friction explicitly, along with solvent viscosity.

In simulations, the shear viscosity of the solvent is calculated from the integral over time of the stress autocorrelation function following the Green-Kubo relation[14]

$$\eta = \frac{1}{Vk_BT}\int_0^\infty dt \left\langle \sigma^{\alpha\beta}(t).\sigma^{\alpha\beta}(0) \right\rangle \tag{18}$$

Here, $V$ denotes the volume of the system, $k_B$ is the Boltzmann constant, $T$ denotes temperature, and $\sigma^{\alpha\beta}$ is the off-diagonal component of the stress tensor, given by

$$\sigma^{\alpha\beta} = \sum_{i=1}^N \left[ m_i v_i^\alpha v_i^\beta + \frac{1}{2}\sum_{\substack{j=1\\j\neq i}}^N F_{ij}^\beta r_{ij}^\alpha \right] \tag{19}$$

In Eq.(19), $m_i$ is the mass, $v_i^\alpha$ and $v_i^\beta$ are the $\alpha$ and $\beta$ components of the velocity of the $i^{th}$ particle, $F_{ij}^\beta$ denotes the $\beta$-component of the force exerted on particle $i$ by particle $j$, and $r_{ij}^\alpha$ represents the $\alpha$-component of the distance separation ($r_{ij}$) between particle $j$ and $i$ (i.e., $r_{ij} = r_j - r_i$). The indices $\alpha, \beta = x, y, z,$ and $\alpha \neq \beta$. Therefore, in three dimensions, the term inside the angular bracket in Eq.(18) becomes

$\sigma^{\alpha\beta}(t)\sigma^{\alpha\beta}(0) = \sigma^{xy}(t)\sigma^{xy}(0) + \sigma^{xz}(t)\sigma^{xz}(0) + \sigma^{yx}(t)\sigma^{yx}(0) + \sigma^{yz}(t)\sigma^{yz}(0) + \sigma^{zx}(t)\sigma^{zx}(0) + \sigma^{zy}(t)\sigma^{zy}(0)$.

Due to its oscillatory behavior, the stress-stress correlation function always suffers a convergence issue in the long-time limit. Therefore, to improve the accuracy of the results, we carry out the averaging over the six stress-auto correlation functions obtained by the six components of the off-diagonal stress tensor, i.e., $\sigma^{xz}, \sigma^{yz}, \sigma^{xy}, \sigma^{yx}, \sigma^{zy},$ and $\sigma^{zx}$. The angular bracket in Eq.(18) denotes the ensemble average. We calculate the shear viscosity ($\eta$) of the LJ and soft sphere systems by employing Eq.(18) and report it in **Table 2**.

**Table-2: The shear viscosity of LJ and Soft sphere systems at a reduced temperature T\*=0.8 and reduced density ρ\*=0.7. The shear viscosity is calculated by employing Eq.(18) for both cases.**

| T\*=0.8,ρ\*=0.7 | LJ | SS |
|---|---|---|
| **Shear viscosity** | 0.89 ±0.2 | 0.74 ±0.2 |

Now, it's time to study the size dependence of Stokes Law friction obtained by pulling with a constant force along a particular direction. We considered only the equal-sized tagged and solvent molecules in the preceding section. Now, we vary the diameter of the tagged particle ($d_{tagged}$), keeping the diameter of the solvent molecules ($d_{sol}$) intact. In this study, we consider the diameter of the tagged particle ($d_{tagged}$) to be 0.5, 1.0, 1.5, and 2.0 times that of the solvent molecules ($d_{sol}$). The Lorentz-Berthelot rule sets the interaction parameters between the tagged particle and solvent molecules since it serves as a necessary approximation in dense liquids dominated by the repulsive part of the intermolecular potential. Several studies have used it earlier, including the soft sphere system.[6,50,50]

In **Figure 3**, we plot the drift velocity of the tagged particle against the applied force for the different sizes of the tagged particle in the case of LJ and soft sphere systems. We employ Stokes Law to calculate friction by fitting the initial linear regime of the plot of drift velocity against force, as shown by the red dashed line. Friction obtained from this calculation is displayed in **Figure 4**.

There are several features that are remarkable in **Figure 3.** First is the linearity in the dependence of Stokes Law friction (SLF) on the size, from a smaller size ($d_{tagged}$=0.5) to a larger size ($d_{tagged}$= 2.0). Second, a deviation sets in for larger sizes which could be for various reasons, but even then, the ratio of frictions for size unity and two differ by a factor close to two, as predicted by hydrodynamics. Third, as before, the soft sphere system demonstrates better adherence to hydrodynamic predictions.

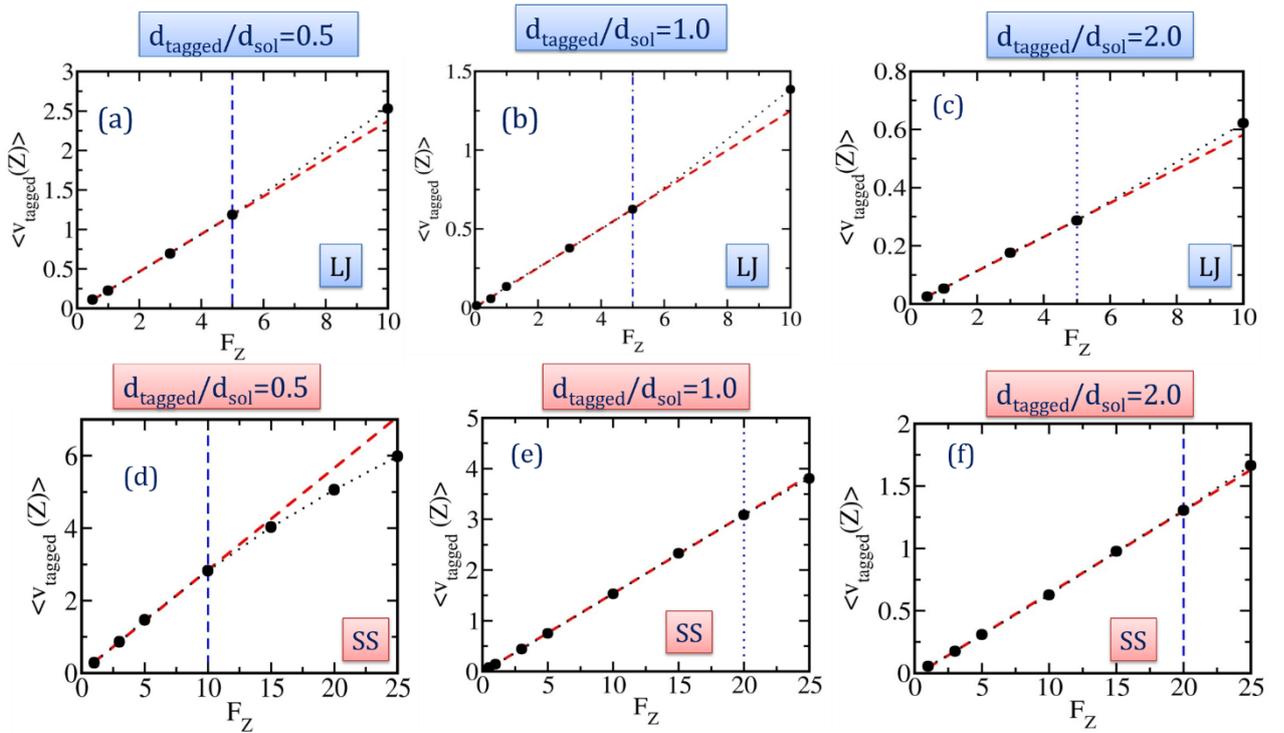

**Figure 3: The plot of the average Z-component steady-state velocity of the tagged particle against the external force for different sizes of the tagged particle in the case of LJ (a-c) and soft sphere (d-f) systems. We vary the diameter of the tagged particle ($d_{tagged}$) from 0.5 to 2.0, keeping the**

**diameter of the solvent molecules unchanged (i.e., d$_{sol}$=1.0), and pull it along the Z-direction by a constant force. In (a), (b), and (c), we show the variation of the Z-component drift velocity against the external force when the ratio of the diameter of the tagged particle and solvent molecule is 0.5, 1.0, and 2.0, respectively, for the LJ system. In all the cases, we observe a deviation from the linearity in the large force region. The linear region (shown by the red dashed line) allows us to estimate friction using Stokes Law. In (d), (e), and (f), we perform the same for the soft sphere (SS) system.**

With the above value of the viscosity, we are in a position to evaluate the numeric constant $C$ that multiplies $\eta a$. In **Figure 4**, we compare the *variation of friction with the size of the tagged particle* obtained from different methods, both in the case of LJ and the soft sphere systems. Stokes Law friction obtained via pulling is plotted against the size ratio, as shown by the black open circles in **Figure 4**. The red-colored diamond in **Figure 4** represents friction obtained from Einstein's relation between diffusion and friction, with the diffusion coefficient obtained directly from the equilibrium simulation in the absence of the external force. We then perform a linear fit to the Stokes Law friction, as shown by the blue dotted line in **Figure 4**. The slope of the linear fit provides an estimate of the numeric constant C with the aid of the viscosity of the solvent. We find a value of $C$ as $5.6\pi$ for LJ and $5.3\pi$ for SS, which is surprisingly close to the Bhattacharyya-Bagchi MCT prediction. [17,18] However, the underlying science is more complex, as discussed in the following sections.

It is observed in **Figure 4** that Stokes friction obtained via pulling exhibits a departure from other methods when the diameter ratio exceeds ~1.5. However, the deviation from theoretical results at large particle sizes deserves special attention. Hydrodynamic relations are expected to become more accurate at large tagged particle sizes.[23] The appearance of a deviation for larger tagged particles is attributed to the pronounced hydrodynamic flow discussed in the subsequent section.[13] We attribute this at least partly to the finite-size effect of the systems simulated. This point deserves further in-depth analysis.

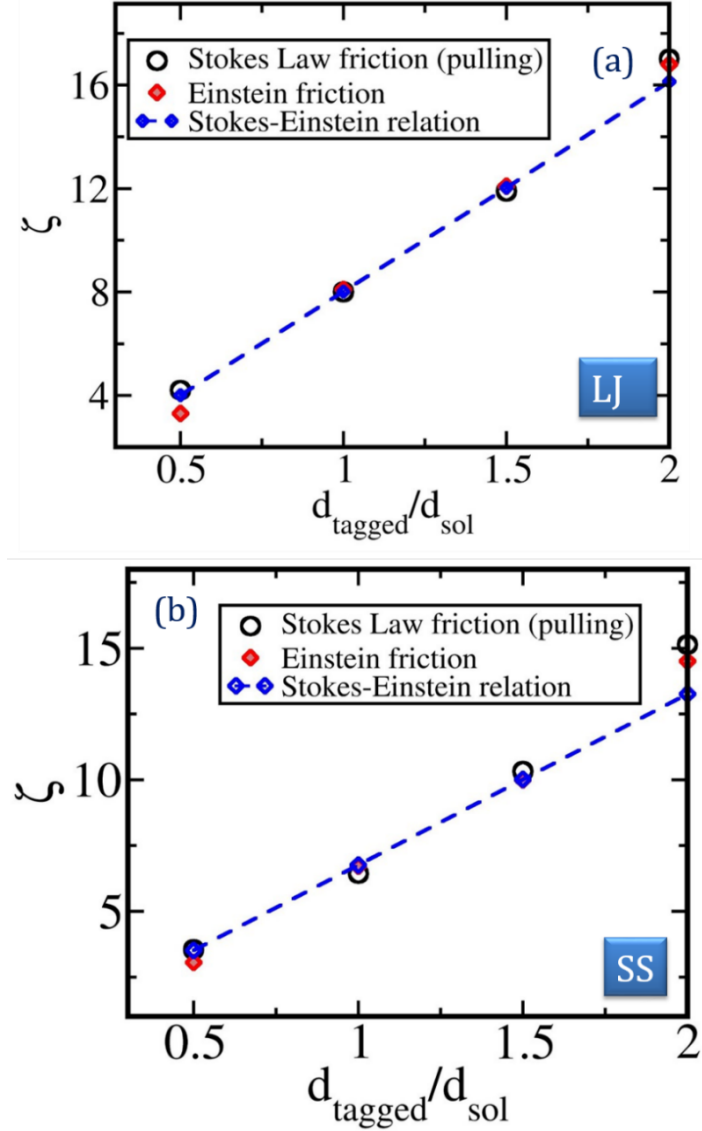

**Figure 4:** Plot of friction against the ratio of the diameter of tagged and solvent particles for the (a) LJ and (b) soft sphere systems. The black open circles denote the friction variation obtained by pulling the tagged particle with a constant force. The red-colored diamonds represent the variation of friction obtained via Einstein's relation, i.e., $D = \dfrac{k_B T}{\zeta}$. We estimate the self-diffusion coefficient $D$ from simulation via mean square displacement and velocity correlation methods. The blue dotted line denotes the variation of friction obtained by employing the Stokes-Einstein relation, i.e., $\zeta = C\eta a$ where $a$ is the radius of the tagged particle. Here, $\eta$ denotes the shear viscosity of the solvent, the value of which is reported in Table 2. Here, $C$ is the numeric constant determined by the hydrodynamic boundary condition. By linear fitting, we find the value of $C$ to be $5.6\pi$ and $5.3\pi$ for LJ and SS, respectively. We observe that friction obtained via different methods is in good agreement with each other up to the diameter ratio (i.e., ($d_{tagged}/d_{sol}$) 1.5.

**However, at the diameter ratio of 2.0, Stokes Law friction obtained via pulling displays a slight deviation from others due to the emergence of pronounced hydrodynamic flow.**

## VI. MICROSCOPIC ASPECTS OF THE SOLVENT RESPONSE

The relationship between force-velocity-friction forms the backbone of the linear response theory in the liquid state. We are pleasantly surprised by two results. First, the linear relation predicted by Stokes Law is indeed valid over an extensive range. Second, the value of the friction thus obtained agrees quantitatively with Einstein friction.

However, the present scheme does break down at large pulling force for all the solute sizes considered. We carry out a detailed study on finite-size effects and find the correction due to the finite size effect can be ignored in our calculations. We discuss this aspect in Supplementary Material (SM-S1). In this context, we perform several analyses to understand the emergence of nonlinear responses beyond the threshold force. We investigate the force spectrum and calculate the smallness parameter at the crossover point in the case of LJ and SS systems (see SM-S2 for the details). We explicitly investigate the velocity profile (SM-S3) and the density profile (SM-S4) of the solvent molecules near the tagged particle along the direction of pulling (Z) and confirm a significant structural deformation near the probe in the relatively large force regime. We also perform a mode-coupling theory based analysis to qualitatively understand the emergence of nonlinear response beyond the threshold force. Our study finds that the reduction of direct collision between the tagged and solvent molecules plays an important role in the deviation from the linearity in the large force regime (see SM-S5 for details).

## VII.  CONCLUSION

As the quantitative relation between an external pulling force, the resultant enforced velocity and the friction acting on the moving sphere given by Stokes' hydrodynamic Law is widely used in experimental, theoretical condensed matter physics, chemistry, and also in time-dependent statistical mechanics, we wanted to explore its quantitative validity. Since the availability of Einstein's relation between diffusion and friction allows us to estimate friction through equilibrium simulations, such a test appears feasible. Stokes' Law is often confused with the Stokes-Einstein relation, which further uses the Stokes relation between friction and viscosity, and this relation depends on the hydrodynamic boundary condition. However, Stokes Law $F^{ex} = \zeta <v_{tagged}>$ discussed here is more general (or less specific). This relation and many of the associated assumptions, including Einstein's theory, were experimentally tested by Perrin using *colloidal particles* in his landmark experiments and were found to agree quantitatively. *However, the validity of Stokes Law for particles of the same size as the solvent/bath molecules appears to have not been tested adequately* in earlier studies and in a similar fashion. While more attention has been focused on the Stokes-Einstein relation that involves viscosity, less attention has been devoted to the original Stokes relation itself.

In the present work, we employ molecular dynamics simulations to probe the validity of Stokes Law for two different interaction potentials. Our approach involves solving the classical Newton's equation of motion for particles. For the solvent particles, we introduce a noise component to conserve the temperature and ensure the establishment of the proper hydrodynamic flow where the solvent velocity varies as $1/r$, with '$r'$ denoting the separation from the center of the tagged particle. In solving the Navier-Stokes (N-S) equation, which yields $\zeta = C\eta a$ viscoelasticity is neglected.[51] This assumption holds true in the limit of a slow time scale, particularly when exerting a constant force on a large colloidal particle. In this study, we demonstrate that this approach yields friction values that quantitatively align with those obtained from diffusion using Einstein's relation across a vast parameter space. Our study also accurately captures the system size dependence, as predicted by theoretical calculations. These two observations, we believe, serve as sufficient assessments regarding the validity of our approach. There are multiple outcomes that are potentially of interest.

(i) First, the existing linear relation between the external pulling force and the resultant steady velocity holds even *at molecular length scales with surprising accuracy*. Additionally, the value of the friction obtained in our pulling experiments works impressively well, as verified by comparing it with Einstein's relation between diffusion and friction, which gives an alternate value of friction.

(ii) While the friction values appear to be not too much different between LJ and SS systems, the breakdown scenario is quite different. The soft sphere system continues to follow Stokes Law until a much (almost four times) higher value of the pulling force than that for LJ system at the studied condition. This is a clear manifestation of the role of intermolecular interactions between the particles at a microscopic level.

(iii) To our surprise, the value of the computed friction is quite close to the prediction of the stick hydrodynamic boundary condition, within ~10%. The present study seems to find a tentative explanation for this in terms of the flow of the particles in the first layer. For the soft-sphere interaction potential, the value of the prefactor "$C$" in the friction-viscosity relation is less than that for the LJ fluid but still closer to the stick boundary condition than the slip.

(iv) However, the present scheme breaks down beyond the threshold force where nonlinear response sets in.

(v) The region of linearity in the plot of drift velocity against the external force is found to be lower at higher liquid density, where particle correlations and cage effects become important. These seem to play a role in limiting the validity of the linear response theory in the context used here for reasons we do not understand at present.

(vi)     We study the velocity profile of solvent around the probe pulled with a constant force along the Z-direction. Along the pulling direction, we find a dramatic change in the behavior of the velocity profile with increasing strength of the force. In order to understand the structural deformation near the probe with increasing strength of the applied force, we investigate the density profile of the solvent along the direction of pulling (Z). Our study shows the existence of an enhanced density in front of the probe along the Z-axis and decreased density in the back -both still lower than the bulk density. We find the emergence of anisotropic behavior becomes more pronounced around the probe beyond the threshold value of the external force.

(vii)    We perform a mode-coupling theory-based analysis to understand the friction on the dragged particle. To understand the binary contribution, we quantitatively calculate Enskog friction for the LJ system with increasing strength of the external force. Our study explicitly shows that the reduction of a direct collision between the tagged particle and the solvent molecules plays a vital role in the breakdown of linear response behavior at large force. The mode coupling theory predicts the increasing importance of the transverse current mode as the density of a liquid is lowered. This aspect deserves further study.

## ACKNOWLEDGEMENTS

We thank Professor Stuart Rice, Professor Roland Netz, and Professor M.L. Klein for fruitful discussions. We thank Dr. Saumyak Mukherjee and Ms. Sangita Mondal for the discussions. This work was supported in part by the National Science Chair Professorship from SERB-DST (India) to BB. SA thanks IISc for the research fellowship.

# Supplementary Material (SM)

## Contents

(S1) Finite-size effects

(S2) Quantifying a smallness parameter for LJ and SS

(S3) Velocity flow profile of the solvent

(S4) Solvent number density profile near the probe

(S5) Mode coupling theory analysis

In this Supplementary Material part, we have presented further analyses with numerical results to support and supplement the main results of the text.

## S1. Finite-size effects

This section quantitatively studies the system size dependence of self-diffusion coefficient and shear viscosity in molecular dynamics simulation under periodic boundary conditions. In a periodic system, the finite size of the simulation cell is expected to influence the results obtained. In this context, Hummer and coworkers derived an analytic expression for finite-size correction based on hydrodynamic arguments[1,2], which has been used in several studies later.[3–5] It is found that for a cubic simulation box of length $L$, the corrected diffusion coefficient ($D_0$) considering the effect of finite size correction is given by [2–4]

$$D_0 = D_{PBC} + 2.83 k_B T / 6\pi \eta L \tag{S1}$$

Here, $D_{PBC}$ is the diffusion coefficient directly obtained from simulation with PBC, $\eta$ is the shear viscosity of the solvent, $k_B$ is the Boltzmann constant, and $T$ is the temperature. Eq.(S1) is useful to correct the observed diffusion coefficient of solutes at infinite dilution in a solvent of known viscosity $\eta$. However, we can calculate the corrected diffusion coefficient even if the shear viscosity of the solvent is not known precisely. In this case, one needs to calculate $D_{PBC}$ for different system sizes.

**Table-S1: We study the variation of friction for five different system sizes. Here, we report the box length of the simulation cell and the total number of particles in the studied system.**

| Total number of particles | Box length |
|---|---|
| 6,912 | 21.45 |
| 10,976 | 25.03 |
| 13,500 | 26.82 |
| 16,384 | 28.60 |
| 23,328 | 32.18 |

Then the corrected diffusion coefficient can be determined from the Y-intercept of a linear fit of $D_{PBC}$ with respect to $1/L$, which corresponds to extrapolation to infinite system size. In **Table S1**, we detail the systems studied in this work. We first compute $D_{PBC}$ for five different sizes of the system. We then use Einstein's relation between diffusion and friction, i.e. $D = \frac{k_B T}{\zeta}$, to calculate friction and plot it against $1/L$, as shown by the blue-colored line in **Figure S1**. It is found that $D$ exhibits a linear dependence on $1/L$, and the intercept of the plot measures the diffusion coefficient corresponding to an infinite-size system. We also determine friction corresponding to that $D$ using Einstein's relation, as shown by the black dotted line in **Figure**

**S1**. At the same time, we estimate Stokes Law friction directly from the simulation for five different system sizes and plot it against *1/L*, as shown by the red dotted line in **Figure S1**.

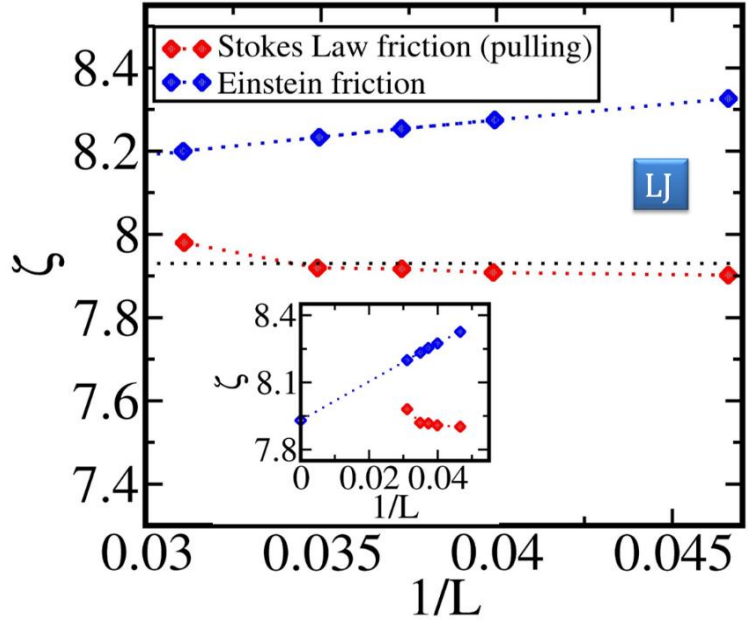

**Figure S1: (a) Plot of friction obtained from different methods against the inverse size (*L*) of the simulation box in the case of the LJ system. The red dotted line shows the variation of Stokes Law friction obtained via pulling with 1/*L*. On the other hand, the blue dotted line shows the variation of friction obtained from Einstein's relation, i.e., $D = \frac{k_B T}{\zeta}$ with D obtained directly from the simulation via MSD. The black dotted line shows the friction variation corresponding to the infinite size system obtained from the intercept. The inset shows the explicit extrapolation of the Einstein friction in the limit.**

**Figure S1** shows that Stokes Law friction obtained via pulling does not exhibit a pronounced linear dependence on 1/*L*, unlike Einstein's friction. Therefore, using Stokes Law, we can ignore the finite size correction in friction calculation without significant error.

## S2. Quantifying a smallness parameter for LJ and SS

In this section, we attempt to quantify the threshold force in terms of the systematic force experienced by the tagged particle in the absence of the external force. **Figure 2** of the main manuscript shows the departure from the linearity in the plot of the drift velocity against the applied force at $F_Z=5.0$ for the LJ and $F_Z=20.0$ for the SS system. We calculate the root mean square fluctuation of the systematic force experienced by the tagged particle, i.e., $\sqrt{\langle \delta F^2_{sys,X/Y/Z} \rangle}$ along the three directions without the external force for LJ and soft sphere systems. We define a quantity $<(\delta F_{sys,Z})^2> = \sqrt{<(F_{sys,Z})^2> - <\overline{F}_{sys,Z}>^2}$ and compute it without the external force. Here $\overline{F}_{sys,Z}$ is the average systematic force experienced by the tagged particle in the absence of external force. Similarly, we calculate the total systematic force experienced by the tagged particle and compute its root mean square fluctuation.

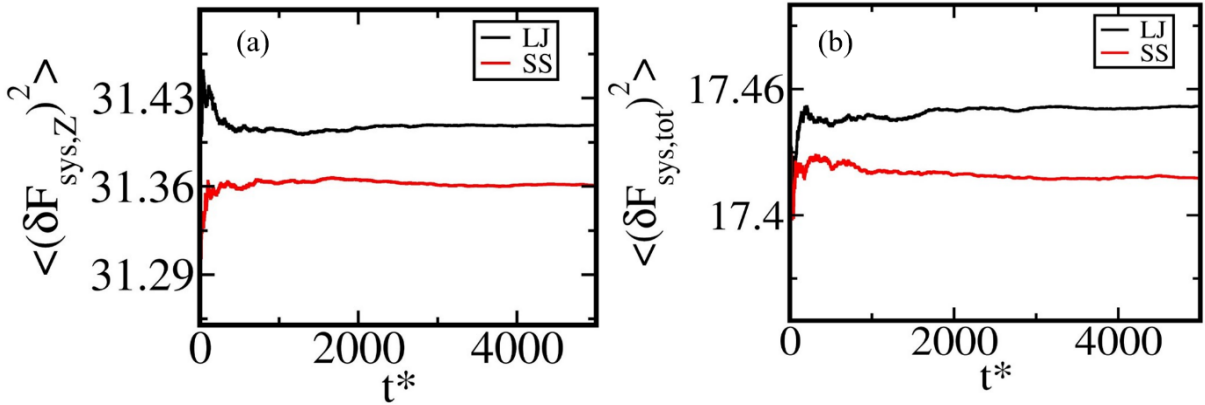

**Figure S2: (a) Plot of the running average of the root mean square fluctuation of the Z-component systematic force experienced by the tagged particle in the absence of the external force against time. Here, the black line denotes the force spectrum for the LJ system, and the red line shows the same for the soft sphere system. (b) We plot the running average of the root mean square fluctuation of the total systematic force experienced by the tagged particle without the external force against time. The black and red lines denote the variation against time for the LJ and SS systems, respectively.**

In **Figures S2a** and **S2b**, we plot the running average of the root mean square fluctuation of the systematic force along the pulling direction and the total force exerted on the tagged particle against time, respectively. To extend our analysis, we introduce the smallness parameter as the ratio of the threshold force (where nonlinear response sets in) and the root mean square fluctuation of the systematic force. We estimate the smallness parameter for LJ and SS systems and report it in **Table S2**. We find that the smallness parameter for LJ and soft sphere systems is significantly different but remains smaller than unity.

**Table-S2: Numerical value of the smallness parameter for LJ and soft sphere systems.**

| Smallness parameter | LJ | SS |
|---|---|---|
| $F_Z / \sqrt{\langle \delta F_{sys,Z/X/Y}^2 \rangle}$ | 0.16 | 0.64 |

Clearly, the smallness parameter for the soft sphere system is ~ four times higher than the same for the LJ system. We perform several additional calculations to understand this discrepancy, as discussed below.

## S3. Velocity flow profile of the solvent

This section studies the behavior of the velocity profile of the solvent around the tagged LJ particle along the direction of pulling. The motivation for the present study comes from the inquiry about the validity of the hydrodynamic boundary condition. *It is challenging to explore and ascertain any signatures of the hydrodynamic boundary condition at molecular length scales.* Note that a stick hydrodynamic boundary condition assumes that the solvent molecules would attach themselves to the surface of the tagged particle and thus move with it. In hydrodynamics, this is the source of energy dissipation.[6,7]

To understand the influence of $<v_{tagged}>$ on the surrounding solvent particles, we divide the spherical layers around the LJ probe into two halves along the *XY*-plane, provided the center of the tagged particle is located at the origin (as shown by the black dotted line in **Figure S3**). We calculate the average Z-component velocities of the solvent molecules in each half of the spherical layers around the LJ probe and plot them against *Z* in **Figure S3**.

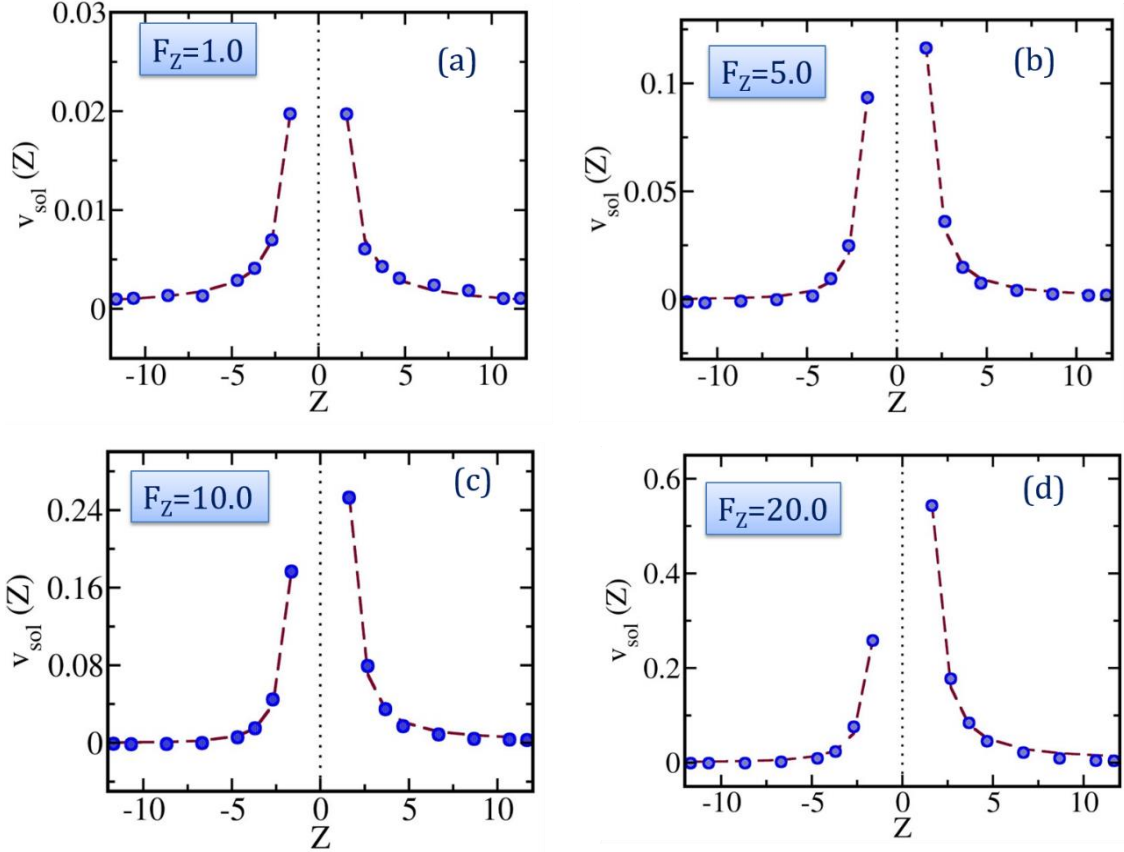

**Figure S3:** Plot of the average Z-component velocity profile of the solvent around the LJ probe pulled with a constant force of (a) $F_Z=1.0$, (b) $F_Z=5.0$, (c) $F_Z=10.0$, and (d) $F_Z=20.0$. We use the expression $v_{sol}(Z) = \dfrac{a_0}{Z} + \dfrac{a_1}{Z^3}$ for fitting the data points, as shown by the deep red-colored lines. Here, $a_0$ and $a_1$ are the fitting constants. We notice the velocity profile to decay at a faster rate than the hydrodynamic prediction. We notice a significant change in the behavior of the velocity profile as we increase the strength of the applied force. The black dotted line indicates the location of the tagged particle.

We observe a noticeable change in the behavior of the velocity profile along the direction of the pulling with the increasing strength of the external force. We observe the velocity profile to exhibit faster decay than the hydrodynamic prediction, which can be interpreted in the frameworks of hydrodynamics theory.[8] *Figure S3 shows that the anisotropic behavior starts to become significant as the strength of the external force exceeds the threshold value, i.e., $F_Z=5.0$ for the LJ system. We study the velocity profile of the solvent for SS in* **Figure**

**S4**. It is found that the soft sphere system exhibits *significant* anisotropic behavior in the velocity profile near the probe around the pulling force $F_Z=20.0$, as shown in **Figure S4**.

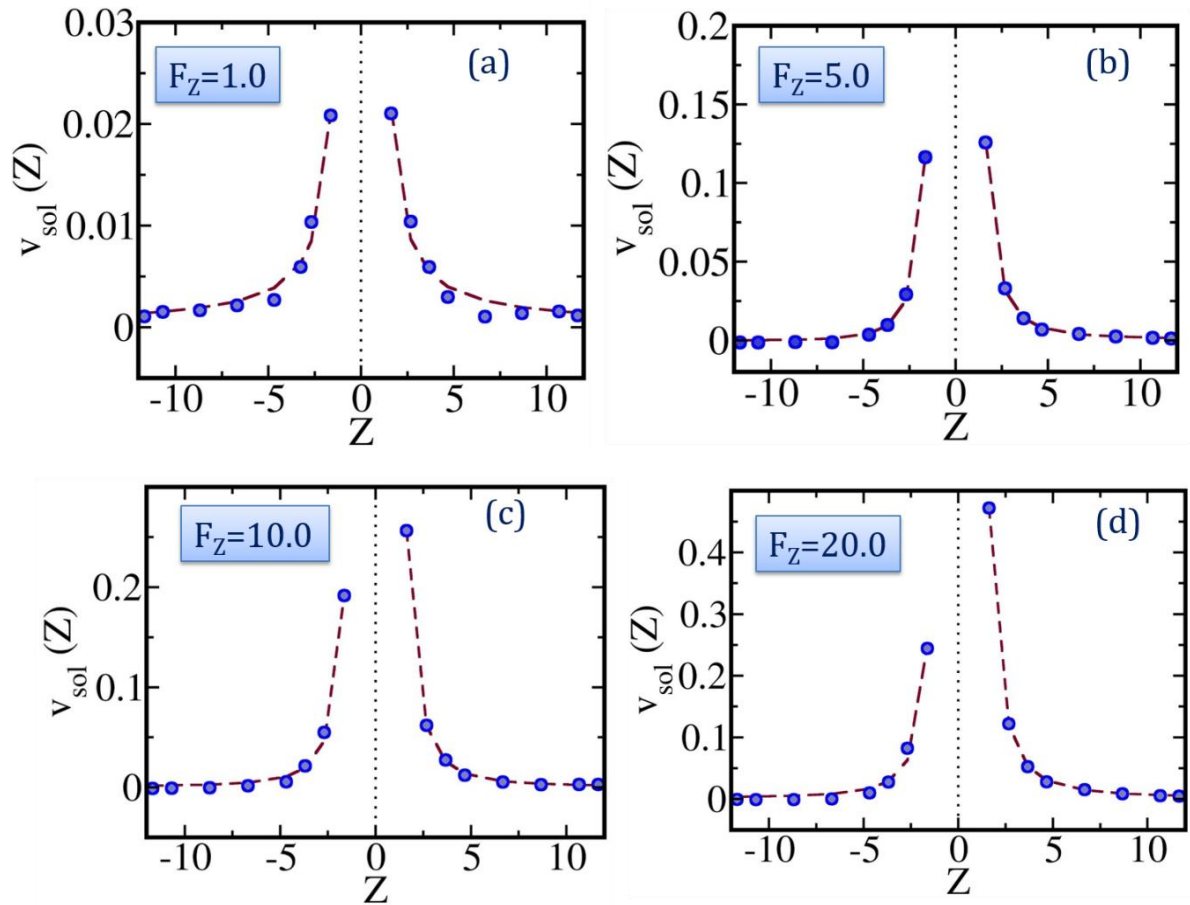

**Figure S4: Plot of the average Z-component velocity of solvent molecules around the SS probe along the direction of pulling (Z) for different values of external force. We use the expression $v_{sol}(Z) = \frac{a_0}{Z} + \frac{a_1}{Z^3}$ for fitting the data points, as shown by the deep red-colored lines. Here, $a_0$ and $a_1$ are the fitting constants. We notice the velocity profile to decay at a faster rate than the hydrodynamic prediction. We observe a significant change in the behavior of the velocity profile as we increase the strength of the applied force. The blue dotted line at Z=0 indicates the location of the tagged particle.**

Apart from the velocity profile along the direction of pulling, we also investigate the velocity profile in the other two perpendicular directions, i.e., Y and X. In the same way, we calculate the velocity profile of the solvent molecules near the probe along the Y direction and plot it against Y in **Figure S5**. The figure shows that the hydrodynamic condition (i.e., $1/r$ dependence with "r" denoting the separation from the center of the tagged particle) is restored in our system. We observe the same behavior along the X-direction as well.

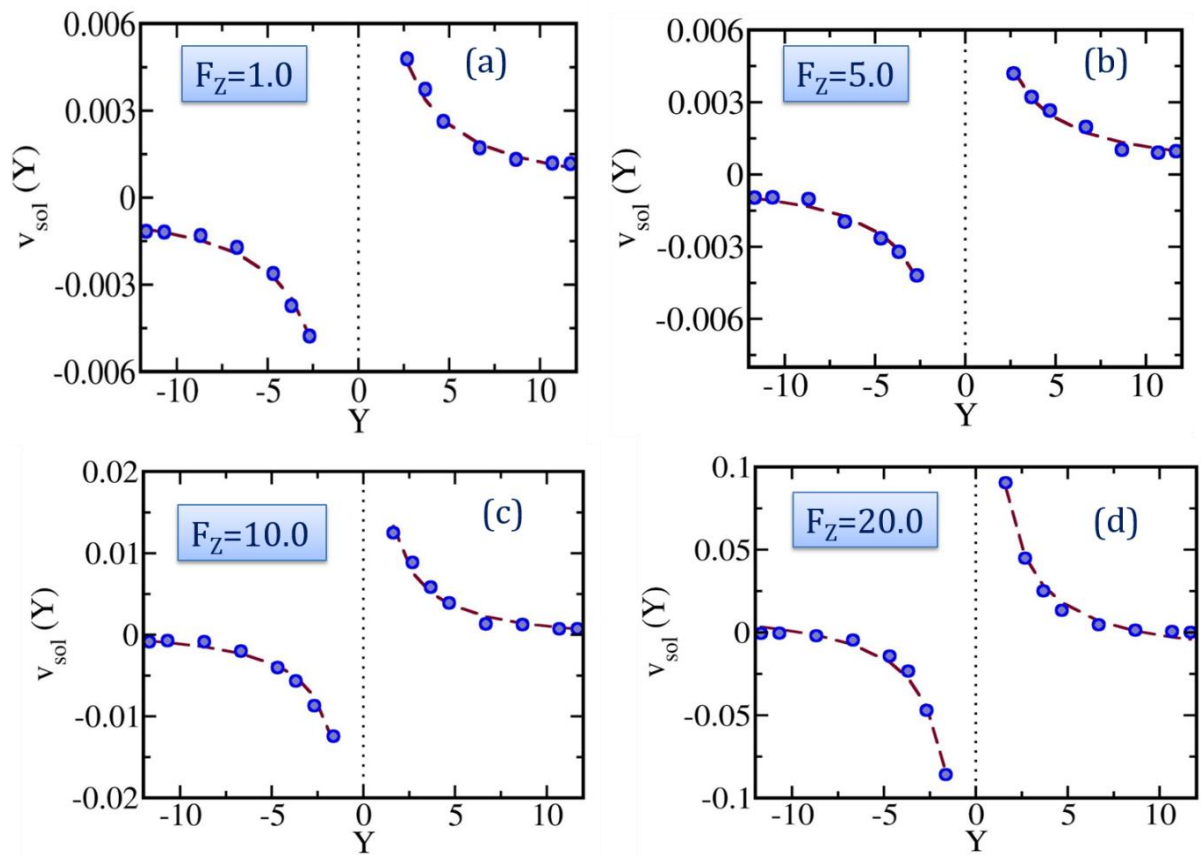

**Figure S5: Plot of the average Y-component velocity of solvent molecules around the LJ probe along the perpendicular direction of pulling (i.e., Y) for different values of external force. We use the expression $v_{sol}(Y) = a_0/Y$ to fit the data points, as shown by the deep red-colored lines. Here, $a_0$ is the fitting constant. We observe the velocity profile to decay following the hydrodynamic prediction. The black dotted line at Y=0 indicates the location of the tagged particle.**

## S4. Solvent number density profile near the probe

In this section, we investigate the emergence of structural deformation near the probe with increasing strength of the applied force. Therefore, we calculate the number density of solvent molecules in the spherical layers around the LJ probe. As discussed earlier, we divide the spherical shell into two halves along the X-Y plane, keeping the tagged particle at the origin, and calculate the number density for each shell along the positive and negative sides of the pulling direction. **Figure S6** plots the number density along the Z-direction for different values of the external force.

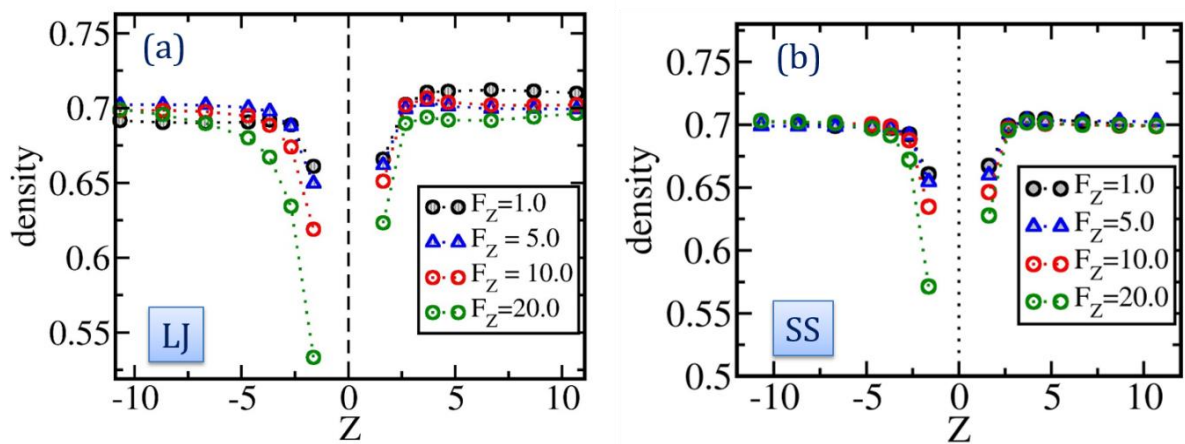

**Figure S6: Plot of the solvent number density profile around the (a) LJ probe and (b) soft sphere probe along the direction of pulling for different values of external force. The black dotted line denotes the location of the tagged particle. It is noted that the tagged particle encounters more solvent molecules on its front with the increasing strength of the external force to resist the motion of the probe.**

At equilibrium, there is a spherically symmetric probability distribution of encountering bath particles, but this symmetry is lost when the tagged particle is pulled with a constant force along the Z-direction. In the vicinity of the dragged sphere, the density differs from the bulk both in front and behind it, exhibiting a decrease in both regions.[9] Notably, the decrease in density is more pronounced behind the moving sphere. This phenomenon can be attributed to the intricate interplay between the entropy-restoring and advective forces.

Brownian motion dissipates the flow energy and removes structural distortions when the shear flow is weak. However, with increasing external force, more solvent molecules per unit of time come in front of the tagged particle to resist the motion of the probe. This is why the number density of the solvent molecules changes significantly on both sides of the tagged particle with increasing force.[10,11]

Due to this anisotropic density distribution, the tagged particle experiences disparate forces on its two sides along the pulling direction. It is essential to highlight that these forces are intimately connected to pressure. Consequently, a pressure disparity arises between the two sides of the tagged particle along the pulling direction. This anisotropy appears perplexing. It is to be noted that the solution of the Navier-Stokes equation indeed predicts more pressure in front, in the limit of low Reynolds number, as discussed below.

The behavior of the density profile in front of the probe confirms a significant structural deformation near the probe in the relatively large force regime ($F_Z$ greater than or equal to 5.0), as evident from **Figure S6**. Notably, the emergence of anisotropic behavior becomes increasingly more pronounced beyond the threshold value for the external force.

## S5. Mode coupling theory analysis

In this section, we employ a mode-coupling theory-based analysis to understand the friction on the dragged particle.[12] According to the standard mode-coupling theory, the frequency-dependent friction on a tagged particle is given by the following exact expression[13,14]

$$\zeta(z) = \frac{1}{k_B T V} \int dr_1 ... dr_2' \left[ \hat{\mathbf{q}} \cdot \nabla_{r_1} v_{12}(\mathbf{r}_1 - \mathbf{r}_2) \right] \times G^s(12;1'2',z) \left[ \hat{\mathbf{q}} \cdot \nabla_{r_1'} v_{12}(\mathbf{r}_1' - \mathbf{r}_2') \right] \quad (S2)$$

Here $G^s(12;1'2',z)$ describes the correlated motion of the tagged particle and the solvent particles. This four-point function represents the time-dependent probability of the tagged particle moving from the position $(r_1', p_1')$ at the time $t'$ to position $(r_1, p_1)$ at time $t$ and a solvent particle which is located at $(r_2', p_2')$ at $t'$ and the same or some other solvent molecule is found at $(r_2, p_2)$. Here $z$ is the Laplace frequency. Due to the separation of time scales between the binary collision and the repeated recollections, Eq.(S2) can be decomposed as [15]

$$\zeta(z) = \zeta_{bin}(z) + \zeta_R(z) \quad \text{(S3)}$$

Where $\zeta_{bin}(z)$ denotes the short-time part of the friction arising due to the direct collision between the solute and solvent particles and $\zeta_R(z)$ is the long-time contribution arising due to the correlated re-collisions of the solute particle with the solvent molecules.

Now, we turn to determine the binary contribution. In our calculation, we are only interested in the zero-frequency value of friction in dense liquids. According to the standard convention, the zero-frequency binary term can be replaced by the Enskog value for friction, which is given by[16,17]

$$\zeta_E = (8/3)\sqrt{2\pi \mu k_B T} \rho \sigma_{12}^2 g(\sigma_{12}) \quad \text{(S4)}$$

Where $\rho$ is the number density of solute, $g(\sigma_{12})$ denotes the value of the radial distribution function at contact, $\mu$ is the reduced mass, and $\sigma_{12} = (\sigma_1 + \sigma_2)/2$ where $\sigma_1$ and $\sigma_2$ are the diameters of solute and solvent molecules, respectively.

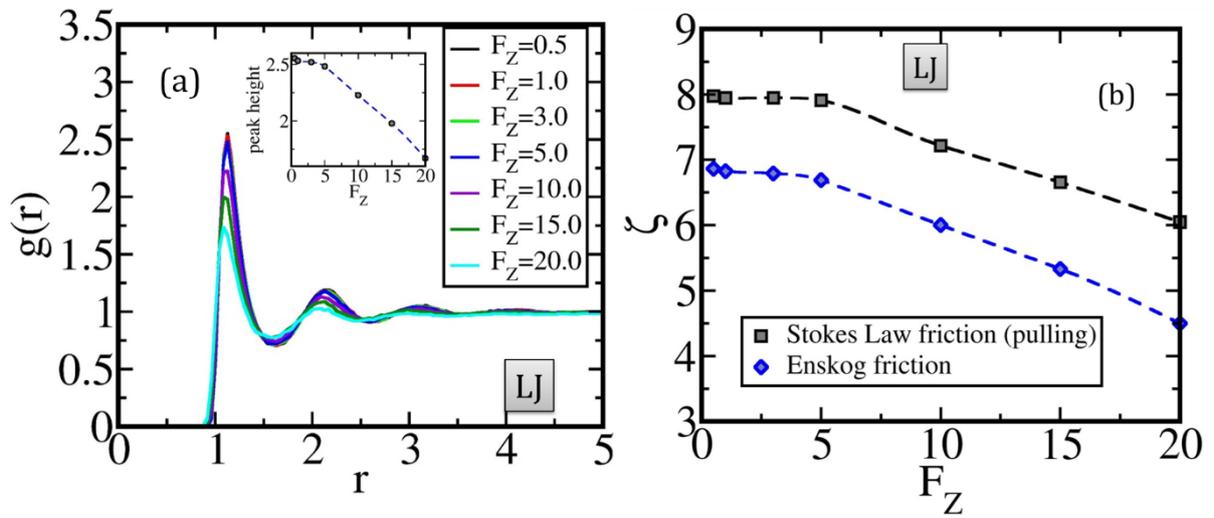

**Figure S7:** (a) Plot of the radial distribution function between the tagged and solvent molecules with increasing strength of the external force. The inset shows the height of the first peak of the radial distribution function with increasing strength of the external force. (b) The plot of the Enskog friction against the external force for the LJ system is shown by the black dotted line. We employ Eq.(S4) to calculate the Enskog friction for different values of external force. Here, the blue dotted line represents the variation of the Enskog friction with the force for LJ system. The black line corresponds to the Stokes Law friction obtained from the simulation via pulling.

In this section, we qualitatively attempt to understand the decrease of Stokes friction beyond the threshold value of the external force using the mode-coupling theory-based analysis in the case of the LJ system. In this regard, we plot the pair distribution function between the probe and solvent molecules for the LJ system with the increasing strength of the external force in **Figure S7 (a)**. The inset shows the height of the first peak. Next, we calculate the Enskog friction by employing Eq.(S4) and plot it against the force for the LJ system, as shown by the blue dotted line in **Figure S7 (b)**. For reference, we show the variation of Stokes Law friction obtained via pulling with the increasing strength of the external force by the black dotted line in **Figure S7 (b)**. It is evident from **Figure S7 (b)** that the Enskog friction pronouncedly decreases with increasing strength of force since the height of the peak of the radial distribution function significantly decreases with force beyond the threshold value. Therefore, the reduction of a direct collision between the tagged and solvent molecules plays a vital role in the deviation from the linearity at a large force.

The calculation of the recollisional term is nontrivial. According to the mode-coupling theory, in the normal liquid regime, Eq.(S3) can be rewritten as [13,15]

$$\frac{1}{\zeta(z)} = \frac{1}{\zeta_{bin}(z) + R_{\rho\rho}(z)} + R_{TT}(z) \tag{S5}$$

Here $R_{\rho\rho}(z)$ is the friction due to the coupling of the solute motion to the collective density mode of the solvent and $R_{TT}(z)$ is the contribution to the diffusion from the current modes of the solvent. If we neglect the contribution from the current modes, then we are left with the term $R_{\rho\rho}(z)$, which is given by

$$R_{\rho\rho}(z) = \frac{\rho k_B T}{m} \int \left[ d\mathbf{k}'/(2\pi)^3 \right] \left( \hat{k}.\hat{k}' \right)^2 k'^2 \left[ c_{12}(k') \right]^2 \times \left[ F^s(k',t)F(k',t) - F_o^s(k',t)F_o(k',t) \right]$$

(S6)

In Eq.(S6), $c_{12}(k)$ is the two-particle direct correlation between the solute and the solvent in the wave number space, $F(k,t)$ is the intermediate scattering function of the solvent, and $F_o(k,t)$ denotes the intertial part of the intermediate scattering function. $F^s(k,t)$ is the self-intermediate scattering function of the solute and $F_o^s$ is the inertial part of $F^s(k,t)$.

Bhattacharyya and Bagchi implemented the above scheme to obtain the friction for the LJ liquid and found a value of friction close to the stick hydrodynamic boundary condition, which was regarded as surprising.[13,15] However, the present Stokes Law and simulation-based work also find a coefficient close to that predicted by the Stokes Law. However, our MCT analysis successfully captures the emergence of nonlinear responses beyond the crossover point.